\documentclass[english,preprint,5p,times,twocolumn]{elsarticle}
\usepackage{amsmath}
\usepackage{graphicx}
\usepackage{amssymb}

\makeatletter

\usepackage{psfrag}
\usepackage{MnSymbol}% the filled square and the circle in caption
\usepackage{babel}
\usepackage{subfloat}
\usepackage{subfig}
\usepackage{esint}

\journal{Journal of Magnetic Resonance}
\makeatother

\begin{document}
\begin{frontmatter}
\title{A mixed basis approach in the SGP-limit}
\author[label1]{Matias Nordin\tnoteref{label3}}

\tnotetext[label3]{Corresponding author. Fax: +46~31160062}
\ead{matias@chalmers.se}

\author[label2]{Martin Nilsson-Jacobi}
\author[label1]{Magnus Nyd\'en}

\address[label1]{Applied Surface Chemistry, Department of Chemical and Biological Engineering, Chalmers University of Technology, 412 96 Gothenburg, Sweden}
\address[label2]{Complex Systems Group, Department of Energy and Environment, Chalmers University of Technology, 412 96 Gothenburg, Sweden}

% use the tnoteref command within \title for footnotes;
%% use the tnotetext command for theassociated footnote;
%% use the fnref command within \author or \address for footnotes;
%% use the fntext command for theassociated footnote;
%% use the corref command within \author for corresponding author footnotes;
%% use the cortext command for theassociated footnote;
%% use the ead command for the email address,
%% and the form \ead[url] for the home page:
%% \title{Title\tnoteref{label1}}
%% \tnotetext[label1]{}
%% \author{Name\corref{cor1}\fnref{label2}}
%% \ead{email address}
%% \ead[url]{home page}
%% \fntext[label2]{}
%% \cortext[cor1]{}
%% \address{Address\fnref{label3}}
%% \fntext[label3]{}

\begin{abstract}
A perturbation method for computing quick estimates of the echo decay in pulsed spin echo gradient NMR diffusion experiments in the short gradient pulse limit is presented. The perturbation basis involves (relatively few) dipole distributions on the boundaries generating a small perturbation matrix in $O(s^2)$ time, where $s$ denotes the number of boundary elements. Several approximate eigenvalues and eigenfunctions to the diffusion operator are retrieved. The method is applied to 1-D and 2-D systems with Neumann boundary conditions.
\end{abstract}
\end{frontmatter}

\section{Introduction}

NMR-methods provide an arsenal of tools to study restricted diffusion~\cite{Grebenkov2007,price2009,callaghan1991}
where not only mass transportal properties such as flow and diffusion
can be studied~\cite{simpson1958,mccall1963,stejskal1965,carr1954}
but also characteristics of the material~\cite{hurlimann1995,sen1995,Price2003,topgaard2003}.
Commonly used for diffusion studies is the pulsed gradient spin-echo
(PGSE) NMR technique where the particle positions are labeled by a
magnetic field gradient~\cite{stejskal1968}. Position labeling is
commonly performed by finite-length magnetic field gradient pulses
and the theory for this experiment is described by the Bloch-Torrey
equations~\cite{torrey1956}. In the short time gradient limit however,
the spin-echo decay simplifies to a Fourier transform over the propagator~\cite{callaghan1991}.
The short time gradient approximation (SGP) is commonly used to describe
the diffusion process when the geometric length scales of the material
are longer that the effective gradient length scales as given in the
q-vector approach~\cite{coy1994a,linse1995,wang1995,mair2002}. In
heterogeneous materials the spin-echo decay normally results in a
function that can be describe by a sum of exponentially decaying functions
resulting in a rather featureless form. However, in structurally well-defined
materials, such as packed mono-disperse micrometer sized beads it can
display detailed features from which material structure details can
be obtained~\cite{callaghan1999,callaghan1992,Coy1994,topgaard2003,coy1994a,Codd2003}.
In addition, in the limit of short gradient pulses, the initial slope
of the spin-echo decay always conveys information of the mean square
displacement independent on material homogeneity/heterogeneity. It
is thus of interest to calculate a spin-echo decay from homogeneous
and heterogeneous materials in order to learn more about the dependence
between material structure and diffusion. The naive approach to calculate the echo decay in the SGP-limit
is done by diagonalizing the diffusion operator.
In this paper we develop a perturbation technique to calculate rough,
but quick estimates of the echo decay, based on approximate eigenfunctions
of the diffusion operator. These approximate eigenfunctions separates
free diffusion and the influence of the material. Interesting features
of the material can thus be analyzed in detail, also for large scale systems.

\section{Theory}

In the short gradient pulse approximation the echo decay is described
by the so called master equation, a Fourier transform over the propagator~\cite{stejskal1965,stejskal1968,callaghan1995}
\begin{equation}
E(q,t)=\frac{1}{V}\langle q|P(r,r',t)|q\rangle\label{eq:master_equation}\end{equation}
 where by the volume term $\frac{1}{V}$ we have assumed that the
initial positions of the particles are equally distributed among the
sample and $\langle q|=e^{i2\pi qz}$, where $q$ is a real and the
applied gradient is in the $z$-direction. The propagator in equation
\ref{eq:master_equation} denotes the ordinary diffusion propagator~\cite{callaghan1995},
which can be expanded in eigenfunction/eigenvalue pairs~\cite{Arfken1995,callaghan1995}
as \[
P=\sum_{i=0}^{\infty}|i\rangle\langle i|e^{-t\lambda_{i}}.\]
 The eigenequation for the eigenfunction/eigenvalue pairs is written
as \begin{equation}
L|i\rangle=\lambda_{i}|i\rangle\label{eq:eigen_eq}\end{equation}
 where $L$ denotes the effective diffusion operator associated with
the boundary conditions, which will be defined in detail below. Note
that the master equation (equation \ref{eq:master_equation}) can
be written as \begin{equation}
E(q,t)=\frac{1}{V}\sum_{i=0}^{\infty}e^{-t\lambda_{i}}|\langle q|i\rangle|^{2}\label{eq:master_eq_eigen_exp}\end{equation}
 i.e. the echo decay is defined by the overlap between incoming modes
$\langle q|$ and the eigenfunctions $|i\rangle$ to the diffusion
operator $L$, weighted by the time-dependent term $e^{-t\lambda_{i}}.$
In general, equation \ref{eq:master_eq_eigen_exp} must be solved
using numerical methods, since the eigenfunctions of the diffusion
operator are known only for simple geometries. We note also that for
periodic boundary conditions the incoming mode $\langle q|$ is an
harmonic function e.g an eigenfunction to $\Delta$ when $q$ is an
integer.  

We write the diffusion operator as \begin{equation}
L=\Delta-S\label{eq:diff_operator}\end{equation}
 where $\Delta$ denotes the ordinary Laplace operator and $S$ denotes
an operator defining the boundary conditions. We will refer to $S$
as the surface operator. We note that the unperturbed problem reduces
to Laplace equation with solutions $\langle q|$ if $q$ is a valid
wave number~\cite{Nordin2009} and it is evident that part of the
perturbation basis need to consist of a set of integer valued $\langle q|$,
for a correct solution in absence of $S$. By the form of equation
\ref{eq:master_eq_eigen_exp} we are motivated to find some set of
functions orthogonal to $\langle q|$ describing the influence of
the surface operator $S$. The eigenfunctions of the diffusion operator
would then be approximated by linear combinations of $\langle q|$
and these unknown vectors. The echo decay in equation \ref{eq:master_eq_eigen_exp}
would then reduce to \begin{equation}
E(q_{i},t)\approx\frac{1}{V}e^{-t\lambda'}|\alpha_{q_i}|^{2}\label{eq:master_eq_expansion}, \end{equation}
 for weights $\alpha_{q_i}$ induced by the surface operator $S$. For simplicity we will omit the index $i$. $\lambda'$
denotes the approximate eigenvalues of $L$. Such a construction will
now be explained in detail.

We construct $S$ by assuming Neumann conditions at the boundary $\Omega$
\begin{equation}
\hat{n}\cdot\nabla\phi(\omega\in\Omega,t)=0\label{eq:bcs}, \end{equation}
 for the (unknown) solution $\phi(r,t)$. The operator $S$ equals
$\hat{n}\cdot\nabla$, and acts as a directional derivative on $\Omega$.
Each eigenfunction of $S$ consist of two $\delta(r-\omega)$-functions
with sign change over $\Omega$ and it is clear that standard perturbation
techniques will not work, as the norm of $S$ is large in the Laplace
basis. By the form of the eigenfunctions to $S$ we will refer to
them as dipoles. Now we Fourier expand the eigenspace of $S$, with
sign change over $\Omega$ to preserve the dipole form and denote
such surface Fourier modes by $\sigma_{s}$. If the surface is smooth,
a Fourier expansion on the boundary captures incoming waves $\langle q|$
of about the same wave numbers. This means that for a truncated set
of Fourier modes $\{\langle q|\}_{q=1}^{N}$ in the low $q$-regime,
a set of low wave-number surface modes suffices. We denote the number
of such surface modes by $M$. The corresponding surface functions
$|s\rangle$, are then calculated by solving Poisson's equation \begin{equation}
|s\rangle=\intop_{\Omega}\frac{1}{|r-\omega|}\sigma_{s}(\omega)d\omega\label{eq:poisson_solution}\end{equation}
where in two dimensions the kernel is replaced by $\log|r-\omega|$.
The approximate solutions to the diffusion problem (equation \ref{eq:eigen_eq})
can then be written as linear combinations of eigenfunctions to the
Laplace operator $\Delta$, $|q\rangle$ and solutions $|s\rangle$
to Poisson's equation (equation \ref{eq:poisson_solution}) \begin{equation}
|i\rangle=\sum_{q}^{N}\alpha_{q}|q\rangle+\sum_{s}^{M}\beta_{s}|s\rangle\label{eq:linear_combination}.\end{equation}
%where the summation index (e.g $q_{n}$) is omitted for simplicity.
This linear combination does not bear sense if $N\rightarrow\infty$,
as of course $\{\langle q|\}_{q=0}^{\infty}$ already form complete
set. If we however restrict ourselves to a subset of eigenfunctions
of $\Delta$, $N<\infty$, the complementary basis spanned by $|s\rangle$
is interesting and proposes a perturbation technique%
\footnote{This restriction also connects naturally with the experimental NMR
setup, where the range of $q$-vectors is not complete but restricted
by the gradient strength available.}. The outline of the mixed basis approach can also be found in\cite{nordin2009arxive}. 

Although the surface distributions $\sigma_{s}(\omega)$ are chosen
to be orthogonal, the corresponding $|s\rangle$ will not be, but
more importantly they nor will be orthogonal to $\langle q|$. The orthogonalization is
however straight forward noting that \[
\langle q|s\rangle=\frac{1}{\lambda_{q}}\langle q|\sigma_{s}\rangle, \]
 since $\Delta$ is self-adjoint. The scalar product between two solutions
to Poisson equation is also involved in the orthogonalization, but
can be treated as follows \begin{multline*}
\langle s|s'\rangle=\int\int\frac{\sigma_{s}(\omega)}{|r-\omega|}d\omega\int\frac{\sigma_{s'}(\omega')}{|r-\omega'|}d\omega'dr\\
=\int\sigma_{s}(\omega)\sigma_{s'}(\omega')\Theta(\omega,\omega')d\omega d\omega'.\end{multline*}
 The interchange of the integration variables is valid provided that
the unit cell is neutral~\cite{ewald1921} and the resulting function
$\Theta(\omega,\omega')$ can be pre-calculated! Importantly the function
$\Theta$ only depends on the unit cell size and number of dimensions
and thus needs only to be calculated once and used as a look-up table.
Furthermore, by noting that both \[
(\Delta-S)|s_{\perp}\rangle\]
 and \[
(\Delta-S)|q\rangle\]
 only need to be calculated over the surface, the resulting perturbation
matrix is quickly accessible and is of size $(N+M)\times(N+M)$. The perturbation matrix has the following form
\[
A_{i,j}=\langle i | (\Delta - S) | j \rangle, 
\]
where $i$ and $j$ range over all perturbation vectors i.e. $\{\langle q |\}_{q=1}^N$ and $\{\langle s_{\perp} |\}_{s=1}^M$. The scalars $\alpha_{q}$ in equation \ref{eq:master_eq_expansion} as well as the approximate eigenvalues can be retrieved by diagonalization
of the small perturbation matrix $A$. We denote the eigenfunctions of
the perturbation matrix by $|k\rangle$. For $t\rightarrow0$, the
echo decay is solely expressed by $|\langle q|k\rangle|^{2}$ and
can directly be read out from the first $N$ elements of the diagonal of the perturbation matrix. Note also that when $t\rightarrow\infty$ the echo decay reduces to the first column of the perturbation matrix.

\section{Results}

The perturbation basis has been validated in several trivial and non-trivial
domains with good results. Three examples are presented here and in
all examples the free space diffusion constant is set to unity.

\begin{figure}

\psfrag{008}[bc][bc]{$E(q,t)$}%
\psfrag{009}[tc][tc]{$qa$}%
%
%% </text>
%
%% <xtick>
%
\def\matlabfragNegXTick{\mathord{\makebox[0pt][r]{$-$}}}
\psfrag{000}[ct][ct]{$0$}%
\psfrag{001}[ct][ct]{$1$}%
\psfrag{002}[ct][ct]{$2$}%
\psfrag{003}[ct][ct]{$3$}%
%
%% </xtick>
%
%% <ytick>
%
\psfrag{004}[rc][rc]{$10^{-3}$}%
\psfrag{005}[rc][rc]{$10^{-2}$}%
\psfrag{006}[rc][rc]{$10^{-1}$}%
\psfrag{007}[rc][rc]{$10^0$}%

\includegraphics{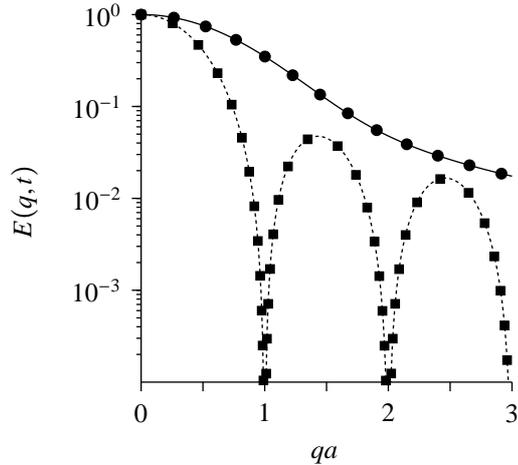}\caption{The figure shows echo decays for diffusion between two plates, separated
by distance $a$. The real echo decay is calculated using equation \ref{eq:master_eq_eigen_exp} with the full
spectrum of the diffusion operator $L$ for time $t=100$ (\textbullet).
The approximate echo decay for the corresponding time is calculated using $N=10$
eigenfunctions to the Laplace operator $\Delta$ and $M=1$ surface
function (solid line). Also shown is the infinite time solution $E(q,\infty)=|\mbox{sinc}(\pi ql)|^{2}$
($\filledmedsquare$) and the approximate infinite time solution (dashed
line) using the same perturbation basis as for the $t=100$ signal.
The approximate signals coincide well with expected results (the relative error is of order $10^{-4}$). \label{fig:echo_decay_plates}}

\end{figure}

The first example consists of diffusion between two plates separated by
a distance $a$, a well studied situation for which an analytic expression
is known~\cite{neuman1974,linse1995,callaghan1995a}. A standard finite
difference approach is used with a grid spacing $h=a/50$. The perturbation
basis consist of $N=10$ eigenfunctions to the Laplace operator
$\Delta$, and one dipole function representing the boundary ($M=1$).
The echo decay is shown in figure \ref{fig:echo_decay_plates} for
$t=100$ and $t=\infty$ together with the real SGP-signal calculated
from the eigenfunction expansion of $L$ and the analytical infinite
time solution $E(q,t\rightarrow\infty)=|\mbox{sinc}(\pi ql)|^{2}$~\cite{price2009}.
The relative error of the approximate echo decay is of order $\sim10^{-4}$
for this perturbation basis and relative error of the apparent diffusion
constant, estimated from the initial slope of the echo decay~\cite{woessner1963},
is also of order $\sim10^{-4}$. 

\begin{figure}
\includegraphics{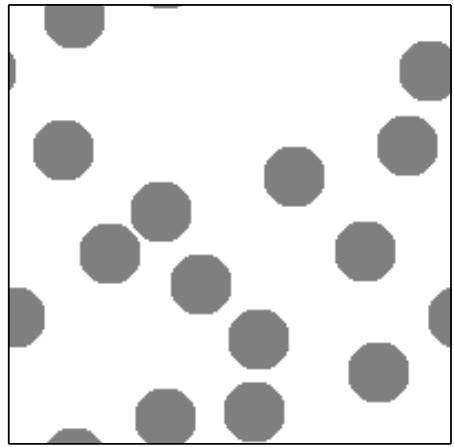}\caption{The figure shows a 2D-system consisting of randomly distributed discs
of equal radius. The system consist of $4\cdot10^{4}$ grid points
and Neumann boundary conditions separates the void space (white region)
from structure (grey region). Figure \ref{fig:discs_echos} shows
the real and approximate echo decay for different times.\label{fig:discs}}

\end{figure}

\begin{figure}

\psfrag{012}[bc][bc]{$E(q,t)$}%
\psfrag{013}[tc][tc]{$ql$}%
%
%% </text>
%
%% <xtick>
%
\def\matlabfragNegXTick{\mathord{\makebox[0pt][r]{$-$}}}
\psfrag{000}[ct][ct]{$0$}%
\psfrag{001}[ct][ct]{$0.1$}%
\psfrag{002}[ct][ct]{$0.2$}%
\psfrag{003}[ct][ct]{$0.3$}%
\psfrag{004}[ct][ct]{$0.4$}%
%
%% </xtick>
%
%% <ytick>
%
\psfrag{005}[rc][rc]{$10^{-6}$}%
\psfrag{006}[rc][rc]{$10^{-5}$}%
\psfrag{007}[rc][rc]{$10^{-4}$}%
\psfrag{008}[rc][rc]{$10^{-3}$}%
\psfrag{009}[rc][rc]{$10^{-2}$}%
\psfrag{010}[rc][rc]{$10^{-1}$}%
\psfrag{011}[rc][rc]{$10^{0}$}%

\includegraphics{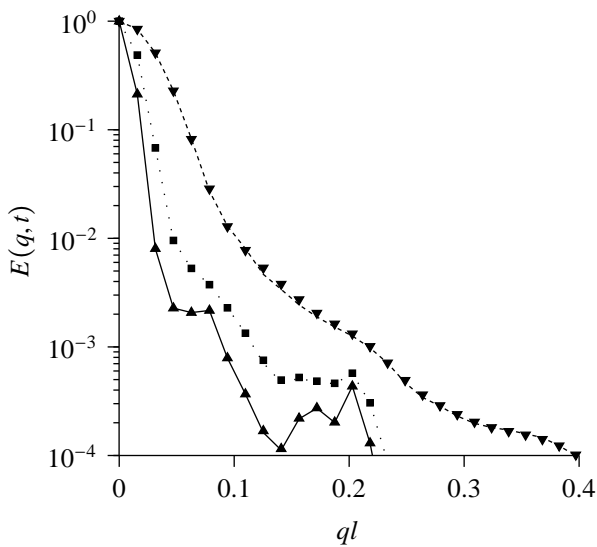}\caption{The figure shows echo decays for the 2D-example of randomly distributed
discs (see figure \ref{fig:discs}). The real echo decay (calculated using equation \ref{eq:master_eq_eigen_exp}) is shown
for times $t=200$ ($\filledmedtriangledown$), $t=900$ ($\filledmedsquare$)
and $t=2000$ ($\filledmedtriangleup$). The approximate echo decay
is calculated using $N=150$ eigenfunctions to the Laplace operator
$\Delta$ and $M=280$ surface functions and is shown for $t=200$
(dashed line), $t=900$ (dotted line) and $t=2000$ (filled line).
The box side length is $l=200$. The times are chosen to approximately double the diffusive length for each time and the relative error of the approximate echo decays is of order $\sim10^{-3}$. Note that the approximative echos are calculated only at the corresponding $q$-values but lines are drawn between these, for visualization.\label{fig:discs_echos}}
\end{figure}

The following two examples consist of two-dimensional systems using $4\cdot10^{4}$ grid points. Periodic
boundary conditions are used on the computational cell, which has a side length $l=200$. Neumann boundary conditions separate the void space (white regions) and the structure (grey regions) and the echo decays are calculated in the void space. The dipole distributions for the boundaries are calculated by diagonalizing the finite difference approximation on the boundary yielding Fourier modes spanning the surface and sign change over the domain preserve the dipole form.
The first such example consist of randomly distributed discs with equal radius (see figure \ref{fig:discs}). Figure \ref{fig:discs_echos} shows the real and approximate echo decays for times $t=200,900,2000$ which are chosen to double the diffusive length in each step. The real echo decay is calculated with equation \ref{eq:master_eq_eigen_exp} using the $280$ first eigenfunctions/eigenvalues to $L$, which gives an error $<10^{-9}$ for $t>200$. The approximate echo decay is calculated using the first $N=150$ eigenfunctions to $\Delta$ and $20$ surface Fourier modes per disc are used, in total $M=280$ surface functions represent the boundaries. The relative error of the echo decay is of order $\sim10^{-3}$. 

\begin{figure*}
\subfloat{

\includegraphics{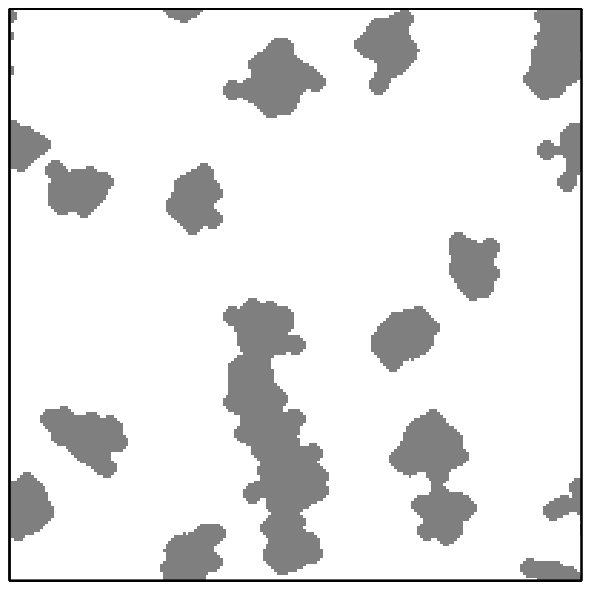}}
\subfloat{
\psfrag{012}[bc][bc]{$E(q,t)$}%
\psfrag{013}[tc][tc]{$ql$}%
%
%% </text>
%
%% <xtick>
%
\def\matlabfragNegXTick{\mathord{\makebox[0pt][r]{$-$}}}
\psfrag{000}[ct][ct]{$0$}%
\psfrag{001}[ct][ct]{$0.1$}%
\psfrag{002}[ct][ct]{$0.2$}%
\psfrag{003}[ct][ct]{$0.3$}%
\psfrag{004}[ct][ct]{$0.4$}%
%
%% </xtick>
%
%% <ytick>
%
\psfrag{005}[rc][rc]{$10^{-6}$}%
\psfrag{006}[rc][rc]{$10^{-5}$}%
\psfrag{007}[rc][rc]{$10^{-4}$}%
\psfrag{008}[rc][rc]{$10^{-3}$}%
\psfrag{009}[rc][rc]{$10^{-2}$}%
\psfrag{010}[rc][rc]{$10^{-1}$}%
\psfrag{011}[rc][rc]{$10^{0}$}%

\includegraphics{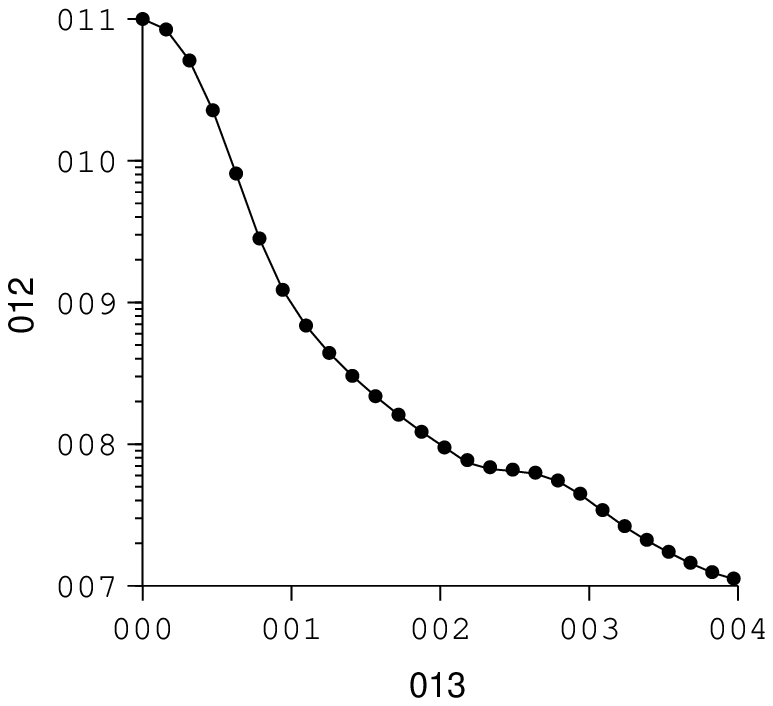}}
\newline
\subfloat{

\psfrag{012}[bc][bc]{$E(q,t)$}%
\psfrag{013}[tc][tc]{$ql$}%
%
%% </text>
%
%% <xtick>
%
\def\matlabfragNegXTick{\mathord{\makebox[0pt][r]{$-$}}}
\psfrag{000}[ct][ct]{$0$}%
\psfrag{001}[ct][ct]{$0.1$}%
\psfrag{002}[ct][ct]{$0.2$}%
\psfrag{003}[ct][ct]{$0.3$}%
\psfrag{004}[ct][ct]{$0.4$}%
%
%% </xtick>
%
%% <ytick>
%
\psfrag{005}[rc][rc]{$10^{-6}$}%
\psfrag{006}[rc][rc]{$10^{-5}$}%
\psfrag{007}[rc][rc]{$10^{-4}$}%
\psfrag{008}[rc][rc]{$10^{-3}$}%
\psfrag{009}[rc][rc]{$10^{-2}$}%
\psfrag{010}[rc][rc]{$10^{-1}$}%
\psfrag{011}[rc][rc]{$10^{0}$}%

\includegraphics{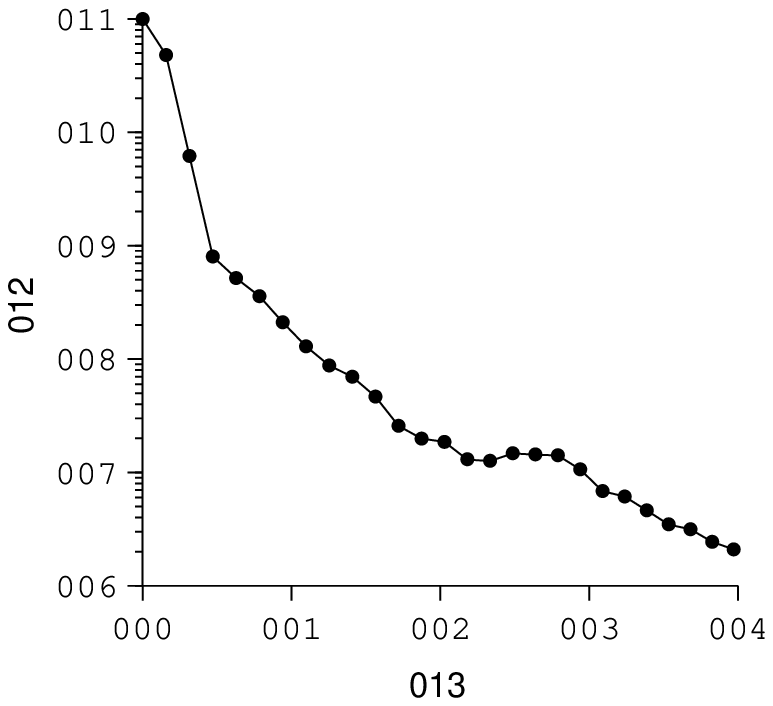}}
%\subfloat{\includegraphics{figures/str5_echost900}}
\subfloat{

\psfrag{012}[bc][bc]{$E(q,t)$}%
\psfrag{013}[tc][tc]{$ql$}%
%
%% </text>
%
%% <xtick>
%
\def\matlabfragNegXTick{\mathord{\makebox[0pt][r]{$-$}}}
\psfrag{000}[ct][ct]{$0$}%
\psfrag{001}[ct][ct]{$0.1$}%
\psfrag{002}[ct][ct]{$0.2$}%
\psfrag{003}[ct][ct]{$0.3$}%
\psfrag{004}[ct][ct]{$0.4$}%
%
%% </xtick>
%
%% <ytick>
%
\psfrag{005}[rc][rc]{$10^{-6}$}%
\psfrag{006}[rc][rc]{$10^{-5}$}%
\psfrag{007}[rc][rc]{$10^{-4}$}%
\psfrag{008}[rc][rc]{$10^{-3}$}%
\psfrag{009}[rc][rc]{$10^{-2}$}%
\psfrag{010}[rc][rc]{$10^{-1}$}%
\psfrag{011}[rc][rc]{$10^{0}$}%

\includegraphics{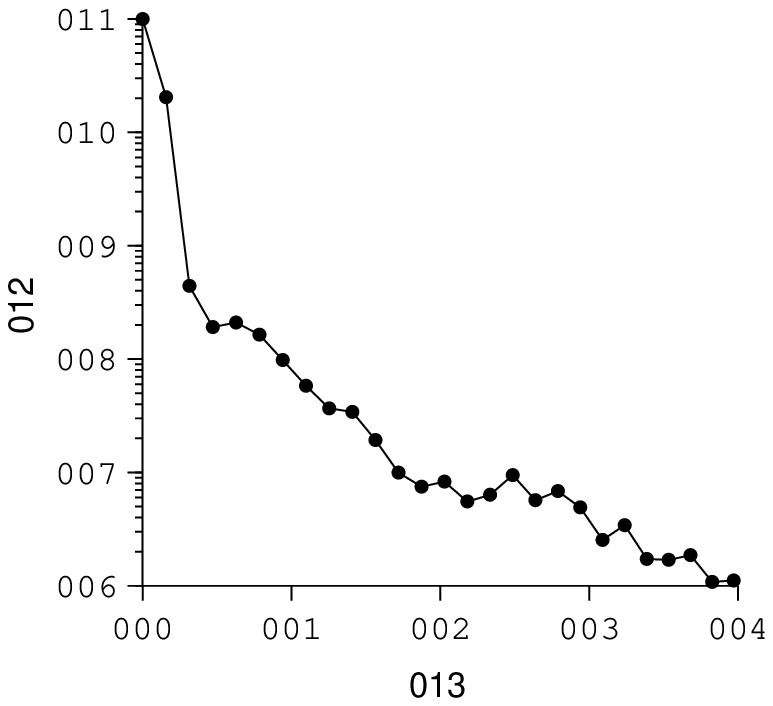}}
\caption{Echo decays for the model shown in top left image for times $t=200$ (top right), $t=900$ (bottom left) and $t=2000$ (bottom right). The times are chosen to approximately double the diffusive length for each step. The real echo decay $E(q,t)$ is calculated using equation \ref{eq:master_eq_eigen_exp} (\textbullet) and the approximative echo decay is calculated using $M=100$ surface functions and the first $N=100$ eigenfunctions of $\Delta$ (solid line). Note that the approximative echo is calculated only at the corresponding $q$-values but the solid line is drawn between these, for visualization. $l=200$ denotes the box side length. The relative error of the approximative echo decay is of order $\sim10^{-2}$. \label{fig:cox_struc}}
\end{figure*}

The last example consists of diffusion in a more interesting 2-dimensional model. The model is generated by a parent/child process~\cite{neyman1958}, where parents are created randomly using a uniform distribution and children are distributed around each parent using a Gaussian distribution. The pixel positions of the children then represent the material. The number of parents/children and the distribution parameters can be varied and the space of geometries is rich. Although such geometries work well in discrete case, they can of course not be spanned by Fourier modes in the continuous limit. Figure \ref{fig:cox_struc} shows an example of one such geometry and echo decays calculated for times $t=200,900,2000$. The perturbation matrix is calculated using the $N=100$ first eigenfunctions to $\Delta$ and $M=100$ surface vectors. In the calculation of the real echo decay the first $230$ eigenfunctions corresponding to the void space are used, which gives an error of order $<10^{-9}$ for $t>200$. The relative error of the approximative echo decay is of order $10^-2$.

\section{Conclusions}
We have shown that for diffusion problems with Neumann boundary conditions
the echo decay in the SGP-limit can be calculated via a perturbation
method with a mixed basis. Approximate echo decays are presented together
with analytic and real echo decays (calculated from the eigenfunction
expansion of the diffusion operator) for trivial and non-trivial
geometries and the relative error of the echo decay is small. The
mixed basis consist of (analytically known) eigenfunctions to the
Laplace operator and solutions to Poisson's equation with dipole distributions
on the boundary. Relatively few base vectors are needed for good result,
resulting in a quickly accessible perturbation matrix. The method is formulated on the boundary, apart from a volume dependent function, which however is geometry independent and can be pre-calculated using standard Ewald summation techniques, saved to disk, and used as a lookup table for arbitrary geometries. This reduces the calculations of approximate propagators and/or echo decays in the SGP-limit to a computational complexity of $O(s^2)$, where $s$ is the number of surface elements. 

As the approximate eigenfunctions not fully compensate for the Neumann
conditions on the boundaries a resonance effect has been observed
when using harmonic functions in the perturbation basis with wave-lengths
corresponding to the structure domains (grey regions), this increases the error of the echo decay at $q$-values corresponding to such wave lengths. At such wave lengths the approximate eigenfunctions consist of linear combinations
of eigenfunctions corresponding to the outer (white region) and inner
(grey regions) domains. This effect can be minimized by increasing
the number of surface modes $M$, preserving the orthogonality to
the inner domains and or not introduce harmonic functions at the
resonance points. Note that the error due to this resonance effect is of the same order as the error at other $q$-values in the geometries presented, but more pronounced.

The method share similarities with
other methods formulated on the boundary such as the boundary element
methods~\cite{banerjee1994} (BEM), analytic element methods~\cite{strack1999}
(AEM) and boundary approximation methods~\cite{li1987,Li2008} (BAE),
also known as Trefftz methods, but might be an alternative due to
the small size of the resulting perturbation matrix achieved in $O(s^{2})$
time where $s$ is the number of boundary elements. The approximate
signals can also be improved by using the approximative eigenfunctions/eigenvalues as an 
initiator for other iterative method as for example~\cite{li2006} which relies on an initial guess of the eigenvalues. The mixed basis approach can also be extended from the SGP-limit to cover time-dependent gradients using the matrix formulation developed by Callaghan~\cite{callaghan1997} based on a multiple propagator approach~\cite{caprihan1996}. As the standard methods for calculations of the diffusion propagator are impractical for large-scale systems, due to the heavy computational demand, the mixed basis approach is suggested as a realistic tool for calculating approximative echo decays, also for finite gradients.

%\bibliographystyle{model1-num-names} 
%\bibliography{references}

\end{document}